\documentclass[12pt,nohyper,notoc]{JHEP}
\usepackage{graphics}
\usepackage{amssymb,epsfig,amsmath,euscript,array,cite}

\newcounter{multieqs}

\newcommand{\bq}{\begin{equation}}
\newcommand{\fq}{\end{equation}}
\newcommand{\bqr}{\begin{eqnarray}}
\newcommand{\fqr}{\end{eqnarray}}

\newcommand{\be}{\begin{equation}}
\newcommand{\ee}{\end{equation}}
\newcommand{\eq}[1]{(\ref{#1})}

\newcommand{\bm}[1]{\mbox{\boldmath $#1$}}

\def\bd{\begin{document}}
\def\ed{\end{document}}
\def\nn{\nonumber}
\def\bea{\begin{eqnarray}}
\def\eea{\end{eqnarray}}
\let\bm=\bibitem
\let\la=\label

\def\npb#1#2#3{Nucl. Phys. {\bf{B#1}} #3 (#2)}
\def\plb#1#2#3{Phys. Lett. {\bf{#1B}} #3 (#2)}
\def\prl#1#2#3{Phys. Rev. Lett. {\bf{#1}} #3 (#2)}
\def\prd#1#2#3{Phys. Rev. {D \bf{#1}} #3 (#2)}
\def\cmp#1#2#3{Comm. Math. Phys. {\bf{#1}} #3 (#2)}
\def\cqg#1#2#3{Class. Quantum Grav. {\bf{#1}} #3 (#2)}
\def\nppsa#1#2#3{Nucl. Phys. B (Proc. Suppl.) {\bf{#1A}}#3 (#2)}
\def\ap#1#2#3{Ann. of Phys. {\bf{#1}} #3 (#2)}
\def\ijmp#1#2#3{Int. J. Mod. Phys. {\bf{A#1}} #3 (#2)}
\def\rmp#1#2#3{Rev. Mod. Phys. {\bf{#1}} #3 (#2)}
\def\mpla#1#2#3{Mod. Phys. Lett. {\bf A#1} #3 (#2)}
\def\jhep#1#2#3{J. High Energy Phys. {\bf #1} #3 (#2)}
\def\atmp#1#2#3{Adv. Theor. Math. Phys. {\bf #1} #3 (#2)}

\newcommand{\EQ}[1]{\begin{equation} #1 \end{equation}}
\newcommand{\AL}[1]{\begin{subequations}\begin{align} #1 \end{align}\end{subequations}}
\newcommand{\SP}[1]{\begin{equation}\begin{split} #1 \end{split}\end{equation}}
\newcommand{\ALAT}[2]{\begin{subequations}\begin{alignat}{#1} #2 \end{alignat}\end{subequations}}
\def\beqa{\begin{eqnarray}} 
\def\eeqa{\end{eqnarray}} 
\def\beq{\begin{equation}} 
\def\eeq{\end{equation}} 

\def\N{{\cal N}}
\def\sst{\scriptscriptstyle}
\def\thetabar{\bar\theta}
\def\Tr{{\rm Tr}}
\def\one{\mbox{1 \kern-.59em {\rm l}}}

\def\a{\alpha}          \def\da{{\dot\alpha}}
\def\b{\beta}           \def\db{{\dot\beta}}
\def\c{\gamma}  \def\C{\Gamma}  \def\cdt{\dot\gamma}
\def\d{\delta}  \def\D{\Delta}  \def\ddt{\dot\delta}
\def\e{\epsilon}                \def\vare{\varepsilon}
\def\f{\phi}    \def\F{\Phi}    \def\vvf{\f}
\def\h{\eta}
\def\k{\kappa}
\def\l{\lambda} \def\L{\Lambda}
\def\m{\mu}     \def\n{\nu}
\def\p{\pi}     \def\P{\Pi}
\def\r{\rho}
\def\s{\sigma}  \def\S{\Sigma}
\def\t{\tau}
\def\th{\theta} \def\Th{\Theta} \def\vth{\vartheta}
\def\X{\Xeta}
\def\z{\zeta}

\def\cA{{\cal A}} \def\cB{{\cal B}} \def\cC{{\cal C}}
\def\cD{{\cal D}} \def\cE{{\cal E}} \def\cF{{\cal F}}
\def\cG{{\cal G}} \def\cH{{\cal H}} \def\cI{{\cal I}}
\def\cJ{{\cal J}} \def\cK{{\cal K}} \def\cL{{\cal L}}
\def\cM{{\cal M}} \def\cN{{\cal N}} \def\cO{{\cal O}}
\def\cP{{\cal P}} \def\cQ{{\cal Q}} \def\cR{{\cal R}}
\def\cS{{\cal S}} \def\cT{{\cal T}} \def\cU{{\cal U}}
\def\cV{{\cal V}} \def\cW{{\cal W}} \def\cX{{\cal X}}
\def\cY{{\cal Y}} \def\cZ{{\cal Z}}

\def\ua{\underline{\alpha}}
\def\ub{\underline{\phantom{\alpha}}\!\!\!\beta}
\def\uc{\underline{\phantom{\alpha}}\!\!\!\gamma}
\def\um{\underline{\mu}}
\def\ud{\underline\delta}
\def\ue{\underline\epsilon}
\def\una{\underline a}\def\unA{\underline A}
\def\unb{\underline b}\def\unB{\underline B}
\def\unc{\underline c}\def\unC{\underline C}
\def\und{\underline d}\def\unD{\underline D}
\def\une{\underline e}\def\unE{\underline E}
\def\unf{\underline{\phantom{e}}\!\!\!\! f}\def\unF{\underline F}
\def\unm{\underline m}\def\unM{\underline M}
\def\unn{\underline n}\def\unN{\underline N}
\def\unp{\underline{\phantom{a}}\!\!\! p}\def\unP{\underline P}
\def\unq{\underline{\phantom{a}}\!\!\! q}
\def\unQ{\underline{\phantom{A}}\!\!\!\! Q}
\def\unH{\underline{H}}

\def\As {{A \hspace{-6.4pt} \slash}\;}
\def\Ds {{D \hspace{-6.4pt} \slash}\;}
\def\ds {{\del \hspace{-6.4pt} \slash}\;}
\def\ss {{\s \hspace{-6.4pt} \slash}\;}
\def\ks {{ k \hspace{-6.4pt} \slash}\;}
\def\ps {{p \hspace{-6.4pt} \slash}\;}
\def\pas {{{p_1} \hspace{-6.4pt} \slash}\;}
\def\pbs {{{p_2} \hspace{-6.4pt} \slash}\;}

\def\Fh{\hat{F}}
\def\Xh{\hat{X}}
\def\ah{\hat{a}}
\def\xh{\hat{x}}
\def\yh{\hat{y}}
\def\ph{\hat{p}}
\def\xih{\hat{\xi}}

\def\psit{\tilde{\psi}}
\def\Psit{\tilde{\Psi}}
\def\tht{\tilde{\th}}
 
\def\At{\tilde{A}}
\def\Qt{\tilde{Q}}
\def\Rt{\tilde{R}}

\def\ft{\tilde{f}}
\def\pt{\tilde{p}}
\def\qt{\tilde{q}}
\def\vt{\tilde{v}}

\def\delb{\bar{\partial}}
\def\bz{\bar{z}}
\def\Db{\bar{D}}

\def\d{\delta}\def\D{\Delta}\def\ddt{\dot\delta}

\def\pa{\partial} \def\del{\partial}
\def\xx{\times}

\def\trp{^{\top}}
\def\inv{^{-1}}
\def\dag{^{\dagger}}
\def\pr{^{\prime}}

\def\rar{\rightarrow}
\def\lar{\leftarrow}
\def\lrar{\leftrightarrow}

\newcommand{\0}{\,\!}      %this is just NOTHING!
\def\one{1\!\!1\,\,}
\def\im{\imath}
\def\jm{\jmath}

\newcommand{\tr}{\mbox{tr}}
\newcommand{\slsh}[1]{/ \!\!\!\! #1}

\def\vac{|0\rangle}
\def\lvac{\langle 0|}

\def\hlf{\frac{1}{2}}
\def\ove#1{\frac{1}{#1}}

\def\Box{\square}
\def\ZZ{\mathbb{Z}}
\def\CC#1{({\bf #1})}
\def\bcomment#1{}
\def\bfhat#1{{\bf \hat{#1}}}
\def\VEV#1{\left\langle #1\right\rangle}

\newcommand{\ex}[1]{{\rm e}^{#1}} \def\ii{{\rm i}}

\title{Dynamical Breaking of Supersymmetry in Noncommutative Gauge Theories }

\author{Chong-Sun Chu$^{a}$, Valentin V. Khoze$^b$ 
and Gabriele Travaglini$^b$\\
$^a$Centre for Particle Theory, Department of Mathematical Sciences,\\
University of Durham, Durham, DH1 3LE, UK\\
$^b$Centre for Particle Theory 
and Institute for Particle Physics Phenomenology,\\
Department of Physics, University of Durham,
Durham, DH1 3LE, UK \\
E-mail: {\tt chong-sun.chu, valya.khoze, 
gabriele.travaglini@durham.ac.uk }}
\abstract{
We propose a new mechanism of spontaneous supersymmetry breaking
in noncommutative gauge theories. We find that in ${\cal N}=1$ noncommutative
gauge theories both supersymmetry and gauge invariance are dynamically broken. 
Supersymmetry is broken spontaneously by a Fayet-Iliopoulos D-term which
naturally arises in a noncommutative $U(n)$ theory. For a non-chiral matter content 
the Fayet-Iliopoulos term is not renormalized and its tree-level value
can be chosen to be much smaller than the relevant string/noncommutativity
scale. In the low energy theory, the noncommutative $U(n)$ gauge symmetry is
broken down to a commutative  $U(1)\times SU(n)$. This breaking is triggered by
the IR/UV mixing and manifests itself at and below the noncommutativity mass
scale $M_{\sst NC}\sim \theta^{-1/2}$.
In particular, the $U(1)$ degrees of freedom decouple from the $SU(n)$ in the
infrared and become arbitrarily weakly coupled, thus playing the role of the
hidden sector for supersymmetry breaking. 
}

\keywords{Non-Commutative Geometry, Supersymmetry Breaking, Gauge Symmetry}

\preprint{{\tt hep-th/0105187}}

\begin{document}

\section{Introduction}

There has been a lot of interest in gauge theories on 
noncommutative spaces. One of the reasons for this interest is the natural
appearance of noncommutativity $[x^\mu, x^\nu]=i\theta^{\mu\nu}$
in the framework of 
string theory and D-branes \cite{CDS,DHull,Chu:1999qz,Schomerus:1999ug,SWnc}.
Noncommutative gauge theories are also fascinating on their own right
mostly due to a mixing between the infrared (IR) and the ultraviolet
(UV) degrees of freedom discovered in \cite{Minwalla}.
This IR/UV mixing does not occur in 
$\N=4$ supersymmetric noncommutative gauge theories \cite{Matusis}.
The $\N=4$ gauge/supergravity correspondence
was analysed in \cite{Hashimoto:1999ut,Maldacena:1999mh}.
The low-energy dynamics of noncommutative $\N=2$
supersymmetric $U(N)$ Yang-Mills theories in the Coulomb phase was recently 
examined in \cite{Armoni:2001br,HKT}, where exact results
were derived for the leading terms in the derivative expansion of the Wilsonian
effective action. In this case the IR/UV mixing is present in the $U(1)$ sector
%cc
\cite{adi,Khoze:2001sy},
but does not affect the $SU(N)$ degrees of freedom. This leads to a dynamical
breaking of noncommutative $U(N)$ gauge symmetry, 
$U(N)\rightarrow U(1)\times SU(N),$ at momentum scales
$k\le M_{\sst NC}\sim \theta^{-1/2},$ with the $U(1)$ degrees of freedom
becoming arbitrarily weakly coupled and approaching a free theory,
$g_{\sst U(1)} \rightarrow 0,$
as $k \rightarrow 0.$ The remaining $SU(N)$ degrees of freedom are
strongly coupled in the IR and are described by the ordinary commutative
Seiberg-Witten solution.
At the same time, in the UV region, 
$k\gg M_{\sst NC}\sim \theta^{-1/2},$ the full noncommutative $U(N)$ gauge
invariance is restored.

In this paper we analyse noncommutative $U(N)$ gauge theories with
$\N=1$ supersymmetry. Our principal result is the observation that these
theories generically exhibit dynamical
supersymmetry breaking (DSB), which does not occur in theories with $\N>1.$

To illustrate this general point we will concentrate here on $\N=1$
noncommutative $U(N)$ theories with fundamental non-chiral matter content such
as SQCD. 
The standard commutative relative
of this theory, $SU(N)$ SQCD, has non-vanishing Witten index
${\cal I}_{\sst SU(N)} =N,$ which is the main topological obstacle for breaking
supersymmetry. However, in the noncommutative set-up the gauge group must 
be $U(N)$ and the Witten index is zero, 
${\cal I}_{\sst U(N)} ={\cal I}_{\sst U(1)}
\cdot{\cal I}_{\sst SU(N)}=0\cdot 1=0.$ Thus, one concludes that supersymmetry
can be at least in principle broken spontaneously even in the non-chiral 
matter context. The simplest scenario of supersymmetry breaking can be
immediately realized by introducing the Fayet-Iliopoulos D-term (FI)
\EQ{L_{\sst FI} =\, \xi_{\sst FI} \int  d^2 \theta d^2 \thetabar
\; \tr_{\sst N} V
\ ,
\label{fiterm}
}
where $V$ is the real $U(N)$ vector superfield\footnote{Superfield
formulation for noncommutative supersymmetric field theories was
introduced in \cite{CZ,susy,tera}. 
}, and the trace over the
$N$ by $N$ matrices selects the $U(1)$-component of $V$.
The FI action, $\int d^4 x L_{\sst FI},$ is $U(N)$ gauge invariant and can be
naturally introduced at tree-level in our $U(N)$ theory. It is well-known
that $\xi_{\sst FI}$ is not renormalized perturbatively or non-perturbatively
\cite{fischler,dine,weinberg} beyond the 1-loop level and the 1-loop correction
trivially vanishes for theories with non-chiral matter. Hence, the tree-level
value of $\xi_{\sst FI}$ is a modulus of the theory and can be taken to be, for
example, $0<\xi_{\sst FI}/ g^2\ll M^2_{\sst NC}.$ 
In this case we will see below
that in our simple SQCD example the supersymmetry breaking scale will be
of order $\sqrt{\xi_{\sst FI}}/ g\ll M_{\sst NC}.$ 
DSB in other theories, including 3-2-1 models, will be analysed elsewhere
\cite{WIP}.

Is noncommutativity really necessary
for this type of supersymmetry breaking? One might consider 
ordinary commutative theories with gauge group $U(1)\times SU(N)$ and introduce
a $U(1)$ FI term as in \cite{dvali3,bidu,ardima}. 
Note however that the $U(1)\times SU(N)$
models can exist only
as low-energy effective theories since in the commutative set-up $U(1)$
will necessarily have a Landau pole in the UV. It is well-known that 
any attempt to grand-unify a $U(1)\times SU(N)$ theory will render the
FI term gauge non-invariant.
On the other hand any unitary {\it noncommutative} gauge theory 
automatically contains the overall $U(1)$ factor which will be
asympotically free in the UV, and will not contain a Landau pole.

We now summarize the results of this paper.
In section 2 we study the supersymmetry breaking pattern in the
$\N=1$ noncommutative SQCD
with gauge group $U(N)$ and $N_{\rm f}< N$ fundamental flavours.
We find that in the presence of a FI term supersymmetry is always spontaneously
broken. We find that the vacuum of the theory lifts the D-flatness condition,
and breaks gauge symmetry down to $U(N-N_{\rm f})$. In section 3 we show that
in the low-energy theory this gauge group manifests itself as a commutative
$U(1) \times SU(N-N_{\rm f})$. We find that the
leading order terms in the derivative expansion of the Wilsonian 
effective action read:
\EQ{
L_{\rm eff}  \ = \ 
-{1  \over 4g^{2}_{\sst 1}(k) } 
 \   F^{\sst U(1)}_{\mu \nu}   
F^{\sst U(1)}_{\mu \nu}    
\ - \ {1  \over 4g^{2}_{\sst N-N_{\rm f}}(k) } 
 \   F^{\sst SU(N-N_{\rm f})}_{\mu \nu}   
F^{\sst SU(N-N_{\rm f})}_{\mu \nu} 
\ + \ 
\cdots
\ , 
\label{uoneres}}
where the dots stand for the superpartners of the 
gauge kinetic terms, terms involving matter fields and higher-derivative
corrections. 
The multiplicative coefficients in front of the gauge kinetic terms in
\eqref{uoneres} define the Wilsonian coupling constants of the corresponding
gauge factors. Their dependence on the Wilsonian scale $k$ is displayed in
Figure 1. 
In particular the running of the $U(1)$ 
has the following asymptotic behaviour:
\beqa
{1\over g^{2}_{\sst 1}(k)}  &\rightarrow&
{3N-N_{\rm f} \over (4\pi)^2} \log k^2 \ , \qquad {\rm as} \ 
k^2\to\infty \ ,\label{rg1}
\\ 
{1\over g^{2}_{\sst 1}(k)}
 &\rightarrow&
{3(N-N_{\rm f}) \over (4\pi)^2} \log {1\over k^2} \ , \qquad {\rm as} \
k^2\to 0 \ , 
\label{rg2}
\eeqa
while that for the $SU(N-N_{\rm f})$ gauge factor is
\beqa
{1\over g^{2}_{\sst N-N_{\rm f}}(k)}  &\rightarrow&
{3N-N_{\rm f} \over (4\pi)^2} \log k^2 \ , \qquad {\rm as} \ 
k^2\to\infty \ ,\label{rg3}
\\ 
{1\over g^{2}_{\sst N-N_{\rm f}}(k)}
 &\rightarrow&
{3(N-N_{\rm f}) \over (4\pi)^2} \log {k^2} \ , \qquad {\rm as} \
k^2\to 0 \ .
\label{rg4}
\eeqa
Notice that the two effective coupling constants are identical in the UV
and run in opposite directions in the IR (below $M_{\sst NC}$).
We interpret this as having full noncommutative $U(N)$ gauge symmetry 
in the UV, which is dynamically broken to $U(1) \times SU(N-N_{\rm f})$
in the low-energy theory.
This intriguing breaking of the gauge symmetry is due to
the IR/UV mixing which affects only the $U(1)$ factor 
%cc
\cite{adi,HKT}.

\begin{figure}[ht]
\begin{center}
{\scalebox{1}{\includegraphics{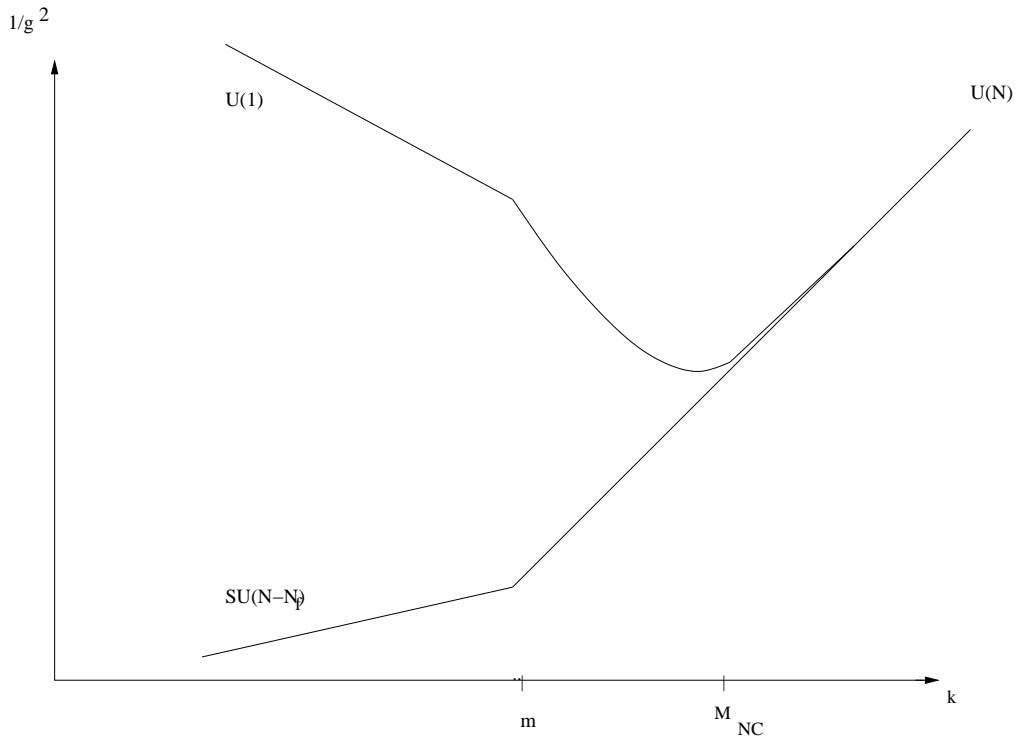}}}
\end{center} \caption{Running of the couplings as a function of the Wilsonian
 scale $k$.  Here $m$ denotes a typical mass of gauge bosons and matter fields.}
\label{cyclic}
\end{figure}

Supersymmetry is broken due to the FI term in the $U(1)$ sector, which 
eventually becomes arbitrarily weakly coupled in the IR.
We thus provide a natural scenario for a gauge-mediated supersymmetry breaking
in which the $U(1)$ factor plays the role of the hidden sector. Both the hidden
sector and the messenger sector are naturally part of the
noncommutative $U(N)$ gauge theory.

\section{Spontaneous Supersymmetry Breaking in Noncommutative SQCD}

Here we will concentrate  on $\N=1$ supersymmetric
noncommutative $U(N)$ QCD with $N_{\rm f}$ flavours $Q$ and $\Qt$.
For concreteness we will consider the case of $N_{\rm f} \le N-1.$
The matter content of the theory is 
described by $N_{\rm f}$
chiral fields $Q^{iI}$ in the representation $\bf{N}$, 
and $N_{\rm f}$ chiral fields $\Qt_{iI}$  in the
anti-fundamental representation $\bar{\bf{N}}$ . Here  
$i =1,\ldots, N; I =1, \ldots, N_{\rm f}$. 
The physical component fields contained in $Q^{iI}$ (resp. $\Qt_{iI}$) 
are  the scalars  $q^{iI}$ and quarks $\psi^{iI}$
(resp.  $\qt_{iI}$ and $\psit_{iI}$).
The noncommutative $U(N)$ gauge symmetry acts
as
\EQ{ \label{un}
Q^{iI} \rightarrow U^i{}_j * Q^{jI}, \quad 
\Qt_{iI} \rightarrow \Qt_{jI}  * (U^{-1})^j{}_i \ ,
}
where $*$ denotes the star-product,
\EQ{(\phi * \chi) (x) \equiv \phi(x) e^{{i\over 2}\theta^{\mu\nu}
\stackrel{\leftarrow}{\partial_\mu}
\stackrel{\rightarrow}{\partial_\nu}}  \chi(x) \ . \label{stardef}}
The $U(N)$ gauge multiplet is
described by the real vector superfield $V=V^A T^A$,
whose physical
components are the vector fields $A_\mu^A$ and the gluinos $\l^A$, 
$A = 0,\ldots ,N^2-1$.  
The field strength superfield is then given by
\EQ{
W_\a = -\frac{1}{4} \Db  \Db e_*^{-2V} * D_\a e_*^{2V} \ .
}
Note that since the star-product only acts in the $x$-space, and
does not affect Grassmann superspace coordinates $\theta$ and $\thetabar$,
the supercovariant derivatives $D$ and $\Db$ behave as constants with respect 
to the star-product.

In the limit of massless flavours, the microscopic Lagrangian 
of SQCD is given by
\EQ{\label{L_0}
L_{\rm micro} = \frac{1}{4 g^2} (\int d^2 \th \; W^A *W^A  + h.c.) + 
\int d^4 \th \; (Q\dag * e_*^{2V} * Q+ \Qt* e_*^{-2V} *\Qt\dag ) \  .
% +  (\int  d^2 \th \cW (Q, \Qt) + h.c. )
} 
The anomaly-free global symmetry of the massless theory is
\EQ{
G = SU(N_{\rm f})_{\rm left}\times SU(N_{\rm f})_{\rm right} 
\times U(1)_R   \ . \label{globsym}
}
Here $U(1)_R$ denotes the anomaly free combination of the axial 
symmetry $U(1)_A$ and the R-symmetry $U(1)_X$. Also note that
the vector symmetry $U(1)_V$ is not included into \eqref{globsym}
as it is a subgroup of the $U(N)$ gauge symmetry.

A tree-level superpotential can be added to the theory 
\EQ{ \label{mass}
\cW_0= \int d^2 \th \sum_{I=1,J}^{N_{\rm f}} m_I{}^J M_J{}^I
}
which introduces bare masses for flavours. Here we defined the
 meson superfield
\EQ{
M_J{}^I=  \Qt_{iJ} * Q^{iI} \ , 
}
 which is gauge invariant\footnote{It 
appears to be challenging to construct a gauge invariant 
baryon operator.}
under the noncommutative $U(N)$ \eq{un}. $M$ 
transforms in the representation $(\bf{N_{\rm f}},\bf{N_{\rm f}})$
of the chiral group $U(N_{\rm f})_{\rm left}\times  U(N_{\rm f})_{\rm right}$. 
Using the global symmetry \eq{globsym}
of the  massless Lagrangian, 
one can diagonalize the mass matrix: $m_I{}^J =
m_I \delta_I{}^J$. 
The nonrenormalization theorems for F-terms
were shown to hold in the noncommutative case as usual \cite{CZ}. Hence the
tree-level superpotential is not renormalized at any order 
in perturbation theory.

For $N_{\rm f} \leq N-1$, in addition to the tree-level superpotential
$\cW_0$ one has to include a nonperturbative
Affleck-Dine-Seiberg (ADS) superpotential \cite{ADS} which is generated dynamically
exactly in the same manner as in the ordinary
commutative case \cite{ADS, cordes, DK},
\EQ{\label{ADS}
\cW_1= \ 
(N-N_{\rm f}) \ 
\L^\frac{3N - N_{\rm f}}{N-N_{\rm f}} (\det{}_* M )^{-\frac{1}{N-N_{\rm f}}} \ , 
}
where the determinant for a $N_{\rm f}\times N_{\rm f}$ matrix $M$ 
is defined by
\EQ{
\det{}_* M = \epsilon^{i_1 \cdots i_{N_{\rm f}}} M_{i_1}{}^{j_1} * \cdots *
M_{i_{N_{\rm f}}}{}^{j_{N_{\rm f}}} \epsilon^{i_1 \cdots i_{N_{\rm f}}} \ .
}
The origin of this potential is similar to that in the
commutative case. The functional form of $\cW_1$ is determined by
gauge invariance under the noncommutative $U(N)$, the
flavour symmetry $SU({N_{\rm f}})_{\rm left}
\times SU({N_{\rm f}})_{\rm right} $ and the fact that it must have R-charge equal to two.

Finally, and most importantly, we add a  Fayet-Iliopoulos D-term \eqref{fiterm}.
As already mentioned in the Introduction,
the FI action is $U(N)$ gauge invariant and can be
naturally introduced at tree-level in our $U(N)$ theory. The well-known result
\cite{fischler,dine,weinberg}
that $\xi_{\sst FI}$ is not renormalized perturbatively or non-perturbatively
beyond the 1-loop level also holds in the noncommutative case.
The 1-loop correction is proportional to the sum of the $U(1)$ matter charges and
trivially vanishes in SQCD, and in general 
for theories with non-chiral matter. This means that a
tree level FI term $\xi_{\sst FI}$ is protected from any quantum corrections,
perturbative or nonperturbative, and is a modulus of the theory.

We can now determine the vacua of  the theory. The
scalar potential is given by
\EQ{
V= V_1 +V_2, \quad \quad 
V_1 = |f|^2 +|\ft|^2, \quad V_2=  \frac{1}{2g^2} D^2 ,
}
where
\EQ{
f_{iI} = \frac{\del}{\del q^{iI}} (\cW_0(q,\qt)+\cW_1(q,\qt)) \ , \quad 
\ft^{iI} =  \frac{\del}{\del \qt_{iI}} (\cW_0(q,\qt)+\cW_1(q,\qt))\ ,\quad 
}
and 
\EQ{
D^A =  \sum_{I=1}^{N_{\rm f}} 
(q\dag_{iI} T^{Ai}{}_j q^{jI} -  \qt_{iI} T^{Ai}{}_j \qt\dag{}^{jI}) 
- \xi_{\sst FI} \;\d^{A0}.
}
Including the mass term \eq{mass}, the ADS superpotential \eq{ADS} and
the FI D-term \eq{fiterm}, one obtains 
\EQ{
f_{iI} = -  \L^\frac{3N - N_{\rm f}}{N-N_{\rm f}}
(\det{}_* M )^{-\frac{1}{N-N_{\rm f}}} \; \frac{1}{q^{iI}} +  m_I
\qt_{iI}  \ , 
}
% & \ft^{iI} = - \frac{b}{N-N_{\rm f}} \L^\frac{3N - N_{\rm f}}{N-N_{\rm f}}
% (\det M )^{-\frac{1}{N-N_{\rm f}}} \; \frac{1}{\qt_{iI}} +  \frac{m}{g^2} q^{iI}.
% \eea
and a similar expression for $\ft^{iI}$.
In order to find the  minimum of $V$, 
we first  apply the global
rotations in the colour and flavour spaces and bring $q, \qt$ to the
form,
\bea
q^{iI} = v_I \delta^{iI}, \quad 
\qt_{iI} = \vt_I \delta^{iI}, \quad & 1 \leq i \leq N_{\rm f}, \nn\\
 q^{iI} = \qt_{iI} =0, \quad & \mbox{otherwise}.
\eea
It is useful to note that
\EQ{
 \sum_{I=1}^{N_{\rm f}}  q^{iI} q\dag_{jI} = 
 \left\{
\begin{array}{ll}
\delta^{ij} |v_i|^2 , & 1 \leq i,j \leq N_{\rm f}, \cr
0,&  \mbox{otherwise},
\end{array}
\right.
}
and $D^A$ simplifies to 
\EQ{
D^A = \sum_{i=1}^{N_{\rm f}} T^{Ai}{}_i ( |v_i|^2 - |\vt_i|^2) - 
\xi_{\sst FI}  \;\d^{A0} \ .
}
Using the completeness relation for the $U(N)$ generators, 
\EQ{
\sum_{A=0}^{N^2-1} T^{Ai}{}_jT^{Ak}{}_l = \frac{1}{2} \delta_{jk}  \delta_{il},
}
one finds that $V_2$ is minimized for $(v_i, \vt_i)$ satisfying
\EQ{\label{v2v2}
 |v_i|^2 - |\vt_i|^2 = \sqrt{\frac{2}{N}} \; \xi_{\sst FI}, \quad
i = 1, \ldots, N_{\rm f},
} 
or $v_i = \vt_i =0$. We reject the latter solution since for vanishing
VEVs the ADS superpotential diverges. 
We also note 
that the right hand side of \eqref{v2v2} does not depend on $i$. Obviously, 
$V$ will attain its global minimum if there are  $(v_i, \vt_i)$ among
\eq{v2v2} that also satisfy
\EQ{\label{ff}
f^{iI} = \ft_{iI} =0.
}
The solution of \eq{ff} is easily determined. It is
\EQ{\label{vv}
v_I \vt_I^*  = \frac{r^2}{m_I} e^{2 \pi i n/N}, \quad n= 1, \ldots, N, 
}
where
\EQ{
r^2 = \L^2 (\prod_{I=1}^{N_{\rm f}} m_I)^\frac{N_{\rm f}}{N} 
\L^{ \frac{N-N_{\rm f}}{N}} \ .
}
Writing $v_I = R_I e^{i \th_I}, v_I = \Rt_I e^{i \tht_I}$,
the vacua of the theory are given by the solutions $(R,\Rt)$ 
to the equations
\EQ{
R_I^2 - \Rt_I^2 = \sqrt{\frac{2}{N}} \xi_{\sst FI}, \quad
R_I \Rt_I =  \frac{r^2}{m_I},
}
and $ \th_I - \tht_I = 2 \pi i n/N$. 
Note that at the minimum one can write
\EQ{
\label{bob}
\frac{1}{ \xi_{\sst FI}} D^A = \sqrt{\frac{2}{N}}  \tr(T^A P_f) -  \delta^{A0},
}
where 
\begin{equation}
P_f = 
\begin{pmatrix} 
\one_{N_{\rm f}\times N_{\rm f}} & 0_{N_{\rm f}\times (N- N_{\rm f})} \cr 
0_{(N- N_{\rm f})\times N_{\rm f}}& \ \ \ 0_{ (N- N_{\rm f})\times(N- N_{\rm f})  }
\end{pmatrix} 
\end{equation}
%\bf{1}_{N_{\rm f}\times N_{\rm f}}$ 
is the unit matrix in the flavour
space. From Eq. \eqref{bob} it is easy to read off which components 
of $D^A$ are lifted.
The minimum value of $V$ can be easily computed,
\EQ{
\langle V\rangle = \frac{1}{2g^2}\xi_{\sst FI}^2  (1 -\frac{N_{\rm f}}{N})
\ , 
}
and it does not depend on $m_I$ and $\L$. 
Since $\langle V\rangle >0$ supersymmetry is spontaneously broken. 
The mass spectrum of the matter field components is affected by the
FI term $\xi_{\sst FI}$. In particular, the mass squares of the matter
scalars are shifted by $\pm\xi_{\sst FI}/g^2$ while the fermion masses
are unaffected in the leading order. This boson-fermion mass non-degeneracy
is the consequence of supersymmetry breaking.

Vacuum expectation values of the scalars also break the $U(N)$ gauge symmetry
down to $U(N-N_{\rm f})$. 
Then $N_{\rm f}^2$ of 
the broken gauge multiplet degrees of freedom 
acquires masses $g\sqrt{R^2+ \Rt^2},$
and the remaining $2N_{\rm f}(N-N_{\rm f})$ ones get masses
$g\sqrt{(R^2+ \Rt^2)/2}$. Here there are no splittings of the superpartner
masses. 
Finally, since the F-flatness (but not the D-flatness) condition is satisfied in
the vacuum, the Goldstino is simply given by $\sum_{A} \langle D^A \rangle \l^A.$

\section{Dynamical Breaking of Gauge Symmetry}

In \cite{Khoze:2001sy,Zanon:2001nq,HKT}, 
the background field perturbation theory
was applied to noncommutative $U(N)$ gauge theories. 
The gauge field $A_\mu$ is decomposed into a background field
$B_\mu$ and a fluctuating quantum field $N_\mu$,
\EQ{ \label{fdec}
A_\mu= B_\mu + N_\mu. 
}
The effective action $S_{\rm eff}[B]$  is obtained by functionally
integrating out all quantum fluctuations. 
The leading term in the derivative expansion of
the effective action takes the form
\EQ{
S_{\rm eff} [ B ] = 2 \int {d^4 k \over (2\pi)^4} B_\mu^{A} (k) B_\nu^{B} (-k)
\; \Pi_{\mu \nu}^{AB} + \cdots
}
Here $\Pi_{\mu \nu}^{AB}$ is the  Wilsonian polarization tensor 
\cite{Khoze:2001sy}, 
 and in supersymmetric theories it has the form
\EQ{ \label{pimunu}
\Pi_{\mu \nu}^{AB} (k) =
\Pi^{AB} (k^2, \tilde{k}^2) (k^2 \delta_{\mu\nu}-k_\mu k_\nu )
\ \ .
} 
Introducing the  matrix of  Wilsonian couplings through the relation
\EQ{ \label{matg}
\left[{1\over g_{\rm eff}^2 (k)}\right]^{AB} \ = \
 { \delta^{AB} \over g_{\rm micro}^2 } \;  \ + \
4 \Pi^{AB} (k^2, \tilde{k}^2) \ ,
}
one can write the effective action as
\EQ{
S_{\rm eff} = - \left[\frac{1}{4g_{eff}^2}\right]^{AB}\int d^4 x \;
F_{\m\n}^A F_{\m\n}^B + \cdots
\label{35}
}
The one-loop 
polarization tensor for the $U(N)$ theory was analysed 
%cc
and it was found in \cite{HKT} that the planar and 
non-planar contributions are given by 
\bea
[\Pi^{AB} ]^{planar} &=&
N \ \delta^{AB} \Pi^{planar}
\ \ ,
\cr \cr
[ \Pi^{AB}  ]^{np} &=& 
N \ \delta^{A0} \delta^{B0}\Pi^{np} 
\ \ ,
\label{pis} 
\eea 
where
\bea
\label{planarsusy2} 
\Pi^{planar} (k^2) &=&  
{2 \over (4\pi )^2 }\left( \sum_{j, {\bf r}} \alpha_{j} \cC(j) C ({\bf r}) \right)
 \left[ {2\over \epsilon} - \gamma_{E}   
- \int_{0}^{1} dx \ \log \frac{x(1-x) k^2}{4\pi \mu^2} \right]  + O(\epsilon),
\ \  \\
\label{nonplanarsusy} 
\Pi^{np}(k^2, \tilde{k}^2) &=& - {4 C({\bf G}) \sum'_{j} \alpha_j \cC(j) \over (4\pi)^2} 
\int_{0}^{1}dx \  
K_0 (\sqrt{x(1-x)} |k| |\tilde{k}|) 
\ \ . 
\eea
Here $\tilde{k}^\mu = \theta^{\mu\nu}k_\nu$,   $j$ is  a spin index and   
$\alpha_{j}$ is equal to $+1$ ($-1$) for ghost (scalar) fields
and to $+1/2$ ($-1/2$) for 
Weyl fermions (gauge fields). Moreover
\EQ{
\cC(j) \equiv \quad 0 \quad
{\rm for\,scalars,}\qquad \frac{1}{2} \quad {\rm for\ Weyl\ fermions,}
\qquad 2 \quad{\rm for\, vectors},
}
and $C({\bf r})$ 
is the Dynkin index for the representation $\bf{r}$.  The sum in 
\eq{planarsusy2} is extended to all fields in the theory, including
ghosts, whereas the sum in \eq{nonplanarsusy} excludes fields in the
(anti)-fundamental representation, which do not contribute to the non-planar diagrams. For the sake of simplicity the expressions in \eqref{planarsusy2}
and \eqref{nonplanarsusy} are written for the case when all the fields 
propagating in the loops are massless. For massive fields these expressions
have to be modified accordingly \cite{Khoze:2001sy,HKT}. 
In particular, when the momentum scale 
$k$ falls below the mass of a particular field, the contribution of
this field should not be included in the summations.

We now examine more closely the Wilsonian polarization function \eq{pis}.
Note that the planar piece is UV divergent and has the gauge
invariant structure $\d^{AB}$.  These terms, as well as the  other
UV divergent terms that arise in 3- and 4-point functions, were examined in
%cc
\cite{adi,bon}. It was found in that \cite{bon} 
these UV divergences can be subtracted in
the traditional way using counter terms, thus proving that the theory is
1-loop renormalizable. 
However,  a closer examination of the nonplanar
contributions shows that the quadratic part of the effective action 
is not gauge invariant  due to the emergence of 
the gauge non-invariant structure $ \d^{A0} \d^{B0}$. 
It was also argued in \cite{zanon} that in the $\cN=4$ gauge theory, the {\it
complete} (all-orders) effective action is gauge invariant. 
% This is achieved by cancellation of the gauge noninvariant terms 
What we see here is that, as far as the low energy effective action 
\eqref{35} is concerned, the
noncommutative gauge symmetry is broken and 
is replaced by the commutative $U(1) \times SU(N)$ gauge
symmetry. We stress that this does not imply that the theory is
inconsistent. By including all the higher derivative terms, one
might hope to
recover the noncommutative gauge symmetry. But our results show that
in the low energy effective theory the
noncommutative gauge symmetry is replaced by the
commutative one. Temporarily neglecting the masses of the fields in the loops,
the running of the Wilsonian coupling constant is
given by \eq{matg}
\EQ{ \label{4512}
\left[{1\over  g_{\rm eff}^2(k^2)}\right]^{AB} =
{ 3N-N_{\rm f}  \over (4\pi)^2} \left(\log{k^2\over \Lambda^2} -2\right) \delta^{AB}
+ {6N \over (4\pi)^2}
\int_{0}^{1}dx\,
K_{0} ( \sqrt{ x(1-x) } |k||\tilde{k}|) \; \d^{A0} \d^{B0}
\ \ .
}
In the UV regime this directly leads to Eqs. \eqref{rg1},
\eqref{rg3}.
In the IR regime one has to decouple the $N_{\rm f}$
flavours of massive matter as well as all the massive gauge multiplets.
This amounts to setting in \eqref{4512} first $N_{\rm f}=0$ and
then replacing $N \rightarrow N-N_{\rm f}.$
This leads to Eqs. \eqref{rg2}, \eqref{rg4}. The general running of the 
Wilsonian couplings is 
represented in Figure 1. The change of slope is due to
a decoupling of massive fields.

\section*{Acknowledgements} 
We would like to thank Gordy Kane for useful comments and 
discussions on future applications \cite{WIP}. We are also grateful
to Nick Dorey, Tim Hollowood, Adrian Signer and Jon Levell for discussions.
This work was partially supported by a PPARC SPG grant.

\ed
\begin{thebibliography}{99}
%\baselineskip 0pt

\bibitem{CDS} 
A.~Connes, M.~R.~Douglas and A.~Schwarz,
``Noncommutative geometry and matrix theory: Compactification on tori,''
JHEP {\bf 9802} (1998) 003, 
{\tt hep-th/9711162}.

\bm{DHull}
M.~R.~Douglas and C.~Hull,
``D-branes and the noncommutative torus,''
JHEP {\bf 9802} (1998) 008, 
{\tt hep-th/9711165}.
%%CITATION = HEP-TH 9711165;%%

\bibitem{Chu:1999qz}
C.~Chu and P.~Ho,
``Noncommutative open string and D-brane,''
Nucl.\ Phys.\ B {\bf 550}  (1999) 151,
{\tt hep-th/9812219}.
%%CITATION = HEP-TH 9812219;%%

\bibitem{Schomerus:1999ug}
V.~Schomerus,
``D-branes and deformation quantization,''
JHEP {\bf 9906} (1999) 030,
{\tt hep-th/9903205}.
%%CITATION = HEP-TH 9903205;%%

\bibitem{SWnc} 
N. Seiberg and E. Witten,
``String theory and noncommutative geometry,''
 JHEP {\bf 9909} (1999) 032,
{\tt hep-th/9908142}.

\bibitem{Minwalla} 
S.~Minwalla, M.~Van Raamsdonk and N.~Seiberg,
``Noncommutative perturbative dynamics,''
JHEP {\bf 0002} (2000) 020,
{\tt hep-th/9912072}.
%%CITATION = HEP-TH 9912072;%%

\bibitem{Matusis} 
A.~Matusis, L.~Susskind and N.~Toumbas, 
``The IR/UV connection in the non-commutative gauge theories,'' 
JHEP  {\bf 0012} (2000) 002, {\tt hep-th/0002075}.


\bibitem{Hashimoto:1999ut}
A.~Hashimoto and N.~Itzhaki,
``Non-commutative Yang-Mills and the AdS/CFT correspondence,''
Phys.\ Lett.\ {\bf B465} (1999) 142,
{\tt hep-th/9907166}.
%%CITATION = HEP-TH 9907166;%%

\bibitem{Maldacena:1999mh}
J.~M.~Maldacena and J.~G.~Russo,
``Large N limit of non-commutative gauge theories,''
JHEP {\bf 9909} (1999) 025,
{\tt hep-th/9908134}.
%%CITATION = HEP-TH 9908134;%%

\bibitem{Armoni:2001br}
A.~Armoni, R.~Minasian and S.~Theisen,
``On non-commutative N=2 super Yang-Mills,''
{\tt hep-th/0102007}.

\bibitem{HKT}
T.~J.~Hollowood, V.~V.~Khoze and G.~Travaglini,
``Exact results in noncommutative N=2 supersymmetric gauge theories,''
{\tt hep-th/0102045}.
%%CITATION = HEP-TH 0102045;%%

%cc
\bibitem{adi}
A.~Armoni,
``Comments on perturbative dynamics of non-commutative Yang-Mills theory,''
Nucl.\ Phys.\ B {\bf 593}, 229 (2001)
[hep-th/0005208].
%%CITATION = HEP-TH 0005208;%%


\bibitem{Khoze:2001sy}
V.~V.~Khoze and G.~Travaglini,
``Wilsonian effective actions and the IR/UV mixing 
in noncommutative  gauge theories,''
JHEP {\bf 0101} (2001) 026,
{\tt hep-th/0011218}.
%%CITATION = HEP-TH 0011218;%%

\bm{CZ}
C.~Chu and F.~Zamora,
``Manifest supersymmetry in non-commutative geometry,''
JHEP {\bf 0002}  (2000) 022,
{\tt hep-th/9912153}.
%%CITATION = HEP-TH 9912153;%%

\bm{susy}
S.~Ferrara and M.~A.~Lledo,
``Some aspects of deformations of supersymmetric field theories,''
JHEP {\bf 0005}  (2000) 008,
{\tt hep-th/0002084}.
%%CITATION = HEP-TH 0002084;%% 

\bm{tera}
S.~Terashima,
``A note on superfields and noncommutative geometry,''
Phys.\ Lett.\ B {\bf 482} (2000) 276,
{\tt hep-th/0002119}.
%%CITATION = HEP-TH 0002119;%%

\bm{fischler}
W.~Fischler, H.~P.~Nilles, J.~Polchinski, S.~Raby and L.~Susskind,
``Vanishing Renormalization Of The D Term In Supersymmetric U(1) Theories,''
Phys.\ Rev.\ Lett.\  {\bf 47} (1981) 757.
%%CITATION = PRLTA,47,757;%%

\bm{dine}
M. Dine,
``Supersymmetry phenomenology (with a broad brush),''
in {\it Fields, Strings, and Duality} TASI 96 lectures, 
{\tt hep-ph/9612389}.
%%CITATION = HEP-PH 9612389;%%

\bm{weinberg} 
S.~Weinberg,
``Non-renormalization theorems in non-renormalizable theories,''
Phys.\ Rev.\ Lett.\  {\bf 80}  (1998) 3702, 
{\tt hep-th/9803099}.
%%CITATION = HEP-TH 9803099;%%

\bm{WIP} Work in progress.

\bibitem{dvali3}
G.~Dvali and A.~Pomarol,
``Anomalous U(1) as a mediator of supersymmetry breaking,''
Phys.\ Rev.\ Lett.\  {\bf 77} (1996) 3728,
{\tt hep-ph/9607383}.
%%CITATION = HEP-PH 9607383;%%

\bibitem{bidu}
P.~Binetruy and E.~Dudas,
``Gaugino condensation and the anomalous U(1),''
Phys.\ Lett.\ B {\bf 389} (1996) 503,
{\tt hep-th/9607172}.
%%CITATION = HEP-TH 9607172;%%

\bibitem{ardima}
N.~Arkani-Hamed, M.~Dine and S.~P.~Martin,
``Dynamical supersymmetry breaking in models with a Green-Schwarz  mechanism,''
Phys.\ Lett.\ B {\bf 431} (1998) 329,
{\tt hep-ph/9803432}.
%%CITATION = HEP-PH 9803432;%%

\bm{ADS} 
I.~Affleck, M.~Dine and N.~Seiberg,
``Dynamical Supersymmetry Breaking In Supersymmetric QCD,''
Nucl.\ Phys.\ B {\bf 241}  (1984) 493.
%%CITATION = NUPHA,B241,493;%%

\bm{cordes} S.~F.~Cordes,
``The Instanton Induced Superpotential In Supersymmetric QCD,''
Nucl.\ Phys.\ B {\bf 273} (1986) 629.
%%CITATION = NUPHA,B273,629;%%bm{cordes} 

\bm{DK}N.~M.~Davies and V.~V.~Khoze,
``On Affleck-Dine-Seiberg superpotential and magnetic monopoles in  supersymmetric QCD,''
JHEP {\bf 0001} (2000) 015,
{\tt hep-th/9911112}.
%%CITATION = HEP-TH 9911112;%%

\bibitem{Zanon:2001nq}
D.~Zanon,
``Noncommutative N = 1,2 super U(N) Yang-Mills: 
UV/IR mixing and  effective action results at one loop,''
Phys.\ Lett.\ B {\bf 502} (2001) 265,
{\tt hep-th/0012009}.

\bm{bon}L.~Bonora and M.~Salizzoni,
``Renormalization of noncommutative U(N) gauge theories,''
Phys.\ Lett.\ B {\bf 504} (2001) 80,
{\tt hep-th/0011088}.
%%CITATION = HEP-TH 0011088;%%

\bm{zanon}M.~Pernici, A.~Santambrogio and D.~Zanon,
``The one-loop effective action of 
noncommutative N = 4 Super Yang-Mills  is gauge invariant,''
Phys.\ Lett.\ B {\bf 504} (2001) 131,
{\tt hep-th/0011140}.
%%CITATION = HEP-TH 0011140;%%



\end{thebibliography}
